\documentclass[12pt]{article}
\usepackage{amsfonts}
\usepackage{bm}
\usepackage{amsmath}
\usepackage{amssymb}
\usepackage{epsfig}
\usepackage[mathscr]{euscript}
\usepackage{hyperref}

\usepackage{slashed}

\newcommand{\bmat}{\left(\begin{array}}
\newcommand{\emat}{\end{array}\right)}

\def\gtrsim{\mathrel{\raise.3ex\hbox{$>$\kern-.75em\lower1ex\hbox{$\sim$}}}}

\def\a{\alpha}

\def\b{\beta}

\def\-{\hphantom{-}}
\def\ov{\overline}
\def\un{\underline}
\def\s2{\frac{1}{\sqrt2}}

\def\mg{m_{3/2}}
\def\mg2{m^2_{3/2}}

\def\Dsl{\,\raise.15ex\hbox{/}\mkern-13.5mu D} 

\def\be{\begin{equation}}
\def\ee{\end{equation}}
\def\bea{\begin{eqnarray}}
\def\eea{\end{eqnarray}}

\newcommand{\nn}{\nonumber}

\begin{document}

\pagestyle{plain}

\makeatletter
\@addtoreset{equation}{section}
\makeatother
\renewcommand{\theequation}{\thesection.\arabic{equation}}
\pagestyle{empty}
\begin{center}
\ \

\vskip .5cm
\LARGE{\LARGE\bf Heterotic Kerr-Schild Double Field Theory and its double Yang-Mills formulation 
 \\[10mm]}
\vskip 0.3cm

\large{Eric Lescano$^1$ and Sourav Roychowdhury$^2$
 \\[6mm]}
 
 {\small  $^1$ Instituto de Astronom\'ia y F\'isica del Espacio (IAFE-UBA-CONICET)\\ [.01 cm]}
{\small\it Ciudad Universitaria, Pabell\'on IAFE, Ciudad de Buenos Aires (1428), Argentina\\ [.2 cm]}

{\small  $^2$ Chennai Mathematical Institute\\ [.01 cm]}
{\small\it SIPCOT IT Park, Siruseri 603 103, India\\ [.2 cm]}

{\small \verb"elescano@iafe.uba.ar,  souravroy@cmi.ac.in"}\\[1cm]

\small{\bf Abstract} \\[0.5cm]\end{center}
We present a formulation of heterotic Double Field Theory (DFT), where the fundamental fields are in $O(D,D)$ representations. The theory is obtained splitting an $O(D,D+K)$ duality invariant DFT. This procedure produces a Green-Schwarz mechanism for the generalized metric, and a fundamental gauge field which transforms as a gauge connection only to leading order. After parametrization, the former induces a non-covariant transformation on the metric tensor, which can be removed considering field redefinitions, and an ordinary Green-Schwarz mechanism on the b-field. Within this framework we explore perturbative properties of heterotic DFT. We use a relaxed version of the generalized Kerr-Schild ansatz (GKSA), where the generalized background metric is perturbed up to quadratic order considering a single null vector and the gauge field is linearly perturbed before parametrization. Finally we compare the dynamics of the gauge field and the generalized metric in order to inspect the behavior of the classical double copy correspondence at the DFT level.

\newpage

\setcounter{page}{1}
\pagestyle{plain}
\renewcommand{\thefootnote}{\arabic{footnote}}
\setcounter{footnote}{0}

\tableofcontents
\newpage

\section{Introduction}
T-duality is an exact symmetry of string theory \cite{Ashoke}. Its low energy limit, or supergravity limit, can be written in a duality invariant way before compactification within the framework of Double Field Theory (DFT) \cite{Siegel} \cite{DFT} \footnote{For reviews we recommend \cite{ReviewDFT} and Section 2 part A of \cite{lecturesSdC}. We follow the notation of the latter.}. Among the different formulations of string theory here we focus on the low energy limit of the heterotic string and its embedded into the double geometry. The canonical way to construct heterotic DFT was given in \cite{HetDFT}. Following the proposal of \cite{MS2}, the standard construction consist of regrouping some of the fundamental fields into a generalized metric. This object is constructed reorganizing the D-dimensional metric tensor, the b-field and a non-Abelian gauge field, i.e., ${\cal H}={\cal H}(g,b,A)$. The generalized metric is a multiplet of the duality group and together with the generalized dilaton are the fundamental fields of heterotic DFT. Moreover, in \cite{FrameDFT} the generalized frame formalism of heterotic DFT was constructed, and the role of the generalized metric is replaced by the generalized version of the vielbein, which has the same field dependence, ${\cal E}={\cal E}(g,b,A)$. Independently of the fundamental fields of the theory, in the standard formulation of heterotic DFT the gauge field and the gauge symmetry only appear upon parametrization.        

On the other hand, there exists certain tension between the generalized metric formulation of DFT and its generalized frame formulation since both schemes are not always equivalent \cite{Tension}. Similarly to what happens in general relativity, the generalized frame contains extra degrees of freedom which give rise to a gauge fixing procedure: some components of this field are fixed and extra conditions appear if/when, for instance, one desires to work considering perturbations around a background. One simple and powerful proposal to explore this tension is the generalized Kerr-Schild ansatz (GKSA). This proposal was introduced in \cite{KL} for ordinary DFT, while the heterotic version of the ansatz was given in \cite{HetKL}. In both cases the ansatz consists of a linear perturbation of a generalized metric/frame background using a pair of generalized null vectors, plus an arbitrary perturbation for the generalized dilaton. When inspecting the form of the ansatz at the supergravity level, the generalized metric formalism produces perturbations around a background metric tensor, $g_{o}$, a background b-field, $b_{o}$ and a background gauge field, $A_{o}$, i.e.,
\bea
\label{Gpert}
g_{\mu \nu} & = & g_{o \mu \nu} - \frac{\kappa}{1+\frac12 \kappa l.\bar{l}} l_{(\mu} \bar{l}_{\nu)} \\ b_{\mu \nu} & = & b_{o \mu \nu} - \frac{\kappa}{1+\frac12 \kappa l.\bar{l}} l_{[\mu} (\bar{l}_{\nu]} - \frac{1}{\sqrt{2}} j_{i} A_{\nu]}{}^{i} ) \\ 
A_{\mu i} & = & A_{o \mu i} + \frac{1}{\sqrt{2}} \frac{\kappa}{1+\frac12 \kappa l.\bar{l}} l_{\mu} j_{i} \, ,
\label{Apert}
\eea
with $\kappa$ an order parameter, $l$ is a null vector and $\bar l$ and $j$ satisfy
\bea
\bar l^2 + j^2 = 0 \, .
\eea
Both $l$ and $\bar l$, as well as $j$ satisfy geodesic equations. These relations are useful to simplify part of the dynamics, and can be easily imposed at the DFT level \cite{HetKL}. Equations (\ref{Gpert})-(\ref{Apert}), plus a perturbation for the background dilaton, integrate the family of theories that can be described using the GKSA. Some of them are the charged black string \cite{blacks} and the charged dilaton black hole \cite{blackh}, both studied in detail in \cite{HetKL} using the generalized metric formalism. As a corollary, the case $j=0$ and $l=\bar l$ describes the subfamily of theories which can be described using the ordinary Kerr-Schild ansatz \cite{KS}.

Contrary to expectation, the family of theories that can be studied using the GKSA in the generalized frame formalism of DFT is slightly smaller than the theories that can be described in the generalized metric formulation, since the gauge fixing condition forces $j=0$, and then both $l$ and $\bar l$ are null vectors. The main goal of this work is two-folded: 
\begin{itemize}
\item On the one hand, we reformulate heterotic DFT in a double Yang-Mills forms, similarly to the proposal \cite{PK}. Our proposal is equivalent to the standard heterotic DFT, and it is constructed writing $O(D,D+K)$ multiplets in terms of $O(D,D)$ multiplets. This kind of formulation is advantageous because it includes a generalized gauge field/symmetry at the DFT level, and the perturbations of this field can be imposed before parametrization, independently of the fundamental fields of the theory. 

\item On the other hand, we relax the form of the GKSA in order to obtain perturbations for the gauged field in the generalized frame formulation of DFT. The relaxed form of the GKSA was firstly used in \cite{Relaxed}, and here we adapt the proposal to use it in heterotic DFT.
\end{itemize}
Furthermore, the previous points allow us to explore the double copy relation at the DFT level: the equation of motion of the generalized metric can be contracted with a generalized killing vector and single/zeroth copy can be explore before parametrization. 

This work is organized as follows: in Section \ref{Review} we review the basic aspects of DFT and the GKSA in the generalized metric/frame formulation. In Section \ref{DYM} we introduce the double Yang-Mills formulation of heterotic DFT, while we explore the relaxed form of the GKSA in Section \ref{RGKSA}. In Section \ref{DCsec} we explore the classical double copy relation at the DFT level. In Section \ref{Dis} we discuss the results of this work and in Section \ref{Conclu} we conclude with a summary. 

\section{Review of heterotic Kerr-Schild Double Field Theory}
\label{Review}
\subsection{Basics of heterotic Double Field Theory}

Heterotic DFT is defined on a double space with coordinates $X^{\cal M}$ which transform under the fundamental representation of the symmetry group $G=O(D,D+K)$, with ${\cal M}=0, \dots, D-1+K$, and $K$ the dimension of the heterotic gauge group.  

The theory is invariant under a global $G$ symmetry which infinitesimally reads
\bea
\delta_G V_{\cal M} = V_{\cal N} h^{\cal N}{}_{\cal M} \, ,
\label{duality}
\eea
where $V_{\cal M}$ is a generic $G$-multiplet, $h \in o(D,D+K)$ is the $G$-parameter. The invariant metric of $G$ is $\eta_{\cal MN} \in G $ and $G$-invariance forces
\bea
h_{\cal MN} = - h_{\cal NM} \, .
\eea
We use $\eta$ and $\eta^{-1}$ in order to lower and raise all the G-indices.

Infinitesimal generalized diffeomorphisms are defined through a generalized Lie derivative, 
\bea
\hat{\cal L}_\xi V_{\cal M} = \xi^{\cal N} \partial_{\cal N} V_{\cal M} + (\partial_{\cal M} \xi^{\cal N} - \partial^{\cal N} \xi_{\cal M}) V_{\cal N} + f_{\cal MNP} \xi^{\cal N} V^{\cal P} + t \partial_{\cal M} \xi^{\cal M} \, ,
\eea 
where $V_{\cal M}$ is an arbitrary generalized tensor, $t$ is a weight constant and $f_{\cal MNP}$ is the generalized version of the structure constants which satisfy
\bea
f_{{\cal MNP}}=f_{[{\cal MNP}]}\, , \qquad f_{[\cal MN}{}^{ \cal R}f_{{\cal P}] R}{}^{\cal Q}=0\, . \label{consf}
\eea
The strong constraint
\bea
\partial_{\cal M} \blacktriangledown \partial^{\cal M} \blacktriangle & = & 0 \, , \\
\partial_{\cal M}(\partial^{\cal M} \blacktriangledown) & = & 0 \, , \\
f^{\cal M N P} \partial_{\cal M} \blacktriangledown & = & 0 \, ,
\eea
is a sufficient condition to ensure the closure of the generalized diffeomorphisms. In the previous expression $\blacktriangledown$ and $\blacktriangle$ are generic fields or parameters in heterotic DFT.

Considering a generic double vector $V_{\cal N}$, the covariant derivative can be defined as
\bea
\nabla_{\cal M} V_{\cal N} = \partial_{\cal M} V_{\cal N} - \Gamma_{\cal MN}{}^{\cal P} V_{\cal P} \, , 
\eea
where $\Gamma_{\cal MNP}$ is a generalized affine connection. 
We demand 
\bea
\nabla_{\cal M}{\cal H}_{\cal NP}&=&0 \, , \quad \nabla_{\cal M}{\cal \eta}_{\cal NP}=0 \, ,  
\eea
and a vanishing generalized torsion 
\bea
\Gamma_{[\cal MNP]}=0 \, ,
\label{torsion}
\eea
in order to determine some projections of the generalized affine connection \cite{RiemannDFT}.

Heterotic DFT is also invariant under a local double Lorentz  ${\cal H}=O(D-1,1)_L\times O(1, D-1+K)_R$ symmetry generated infinitesimally by a generalized parameter $\Gamma_{\cal AB}$ where ${\cal A}=(\underline{\cal A},\overline {\cal A})$  splitting into $O(D-1,1)_L$  and $O(1,D-1+K)_R$ vector indices, $\underline{\cal A}={\un A}=0,\dots , D-1$  and   $\overline A=(\ov A, \ov i)=0,\dots , D-1+K$, {\textit i.e.}, 
\bea
\delta_{\cal H} V_{\cal A} = V_{\cal B} \Gamma^{\cal B}{}_{\cal A} \, ,
\label{Lambda}
\eea
for a generic ${\cal H}$-vector. The $\cal H$-invariance of $\eta_{\cal AB}$ forces $\Gamma_{\cal AB} = - \Gamma_{\cal BA} \,$.

The fundamental fields of the theory consist of a generalized frame ${\cal E}_{\cal M}{}^{\cal A}$ and a generalized dilaton field $d$. The frame-formulation of DFT demands the existence of two constant, symmetric and invertible ${\cal H}$-invariant metrics $\eta_{{\cal AB}}$ and ${\cal H}_{{\cal AB}}$. The former is used to raise and lower the indices that are rotated by ${\cal H}$ and the latter is constrained according to 
\bea
{\cal H}_{\cal A}{}^{\cal C}{\cal H}_{\cal C}{}^{\cal B} = \delta_{\cal A}^{\cal B}\, .
\eea 
The generalized frame is constrained to relate the metrics $\eta_{{\cal AB}}$ and $\eta_{ {\cal MN}}$ and defines a generalized metric ${\cal H}_{\cal MN}$ from ${\cal H}_{\cal AB}$
\be
\eta_{\cal AB} = {\cal E}^{\cal M}{}_{\cal A}\eta_{\cal MN}{\cal E}^{\cal N}{}_{\cal B}\, , \quad {\cal H}_{\cal MN} = {\cal E}_{\cal M}{}^{\cal A}{\cal H}_{\cal AB} {\cal E}_{\cal N}{}^{\cal B} \, .
\ee
${\cal H}_{\cal MN}$ is also an element of $O(D,D+K)$, \textit{i.e.} 
\be
{\cal H}_{\cal MP}\eta^{\cal PQ}{\cal H}_{\cal QN} = \eta_{\cal MN} \, . \label{GMconstraint}
\ee
The DFT curved projectors are defined as
\bea
{\cal P}_{\cal MN} = \frac{1}{2}\left(\eta_{\cal MN} - {\cal H}_{\cal MN}\right)  \ \ {\rm and} \ \
\ov{\cal P}_{\cal MN} = \frac{1}{2}\left(\eta_{\cal MN} + {\cal H}_{\cal MN}\right)\ ,
\eea
and the same can be done with $\eta_{\cal AB}$ and ${\cal H}_{\cal AB}$ to define ${\cal P}_{AB}$ , $\ov{\cal P}_{{\cal AB}}$. We use the convention that the projectors and their inverses lower and raise  projected indices. 

The flat covariant derivative acting on a generic vector $V_{\cal B}$ is
\bea
{\cal D}_{\cal A} V_{\cal B} = {\cal E}_{\cal A} V_{\cal B} + {\omega}_{\cal AB}{}^{\cal C} V_{\cal C} \, , 
\eea
where ${\cal E}_{\cal A}= \sqrt{2} {\cal E}^{\cal M}{}_{\cal A} \partial_{\cal M}$ and  ${\omega}_{\cal AB}{}^{\cal C}$ is the generalized spin connection which satisfies
\be
{\omega}_{\cal ABC} = - {\omega}_{\cal ACB}\, \qquad \mathrm{and} \qquad {\omega}_{\cal A\overline B \underline C} = {\omega}_{\cal A\underline B \overline C} = 0 \, .
\ee

Unlike general relativity, DFT consists of a generalized notion of geometry and there are not enough compatibility conditions to fully determine a generalized spin connection, $\omega_{\cal ABC}$. Only the totally antisymmetric  and trace parts of $\omega_{\cal ABC}$ can be determined in terms of ${\cal E}_{\cal M}{}^{\cal A}$  and $d$,  i.e.
\be
\omega_{\cal[ABC]} = - {\cal E}_{\cal[A} {\cal E}^{\cal N}{}_{\cal B} {\cal E}_{\cal NC]} - \frac{\sqrt{2}}{3}f_{\cal MNP} {\cal E}^{\cal M}{}_{\cal A} {\cal E}^{\cal N}{}_{\cal B} {\cal E}^{\cal P}{}_{\cal C}\equiv -\frac13{\cal F}_{\cal ABC}\, ,
\label{gralspinconnectionE}
\ee
\be
\omega_{\cal BA}{}^{\cal B} = - \sqrt{2}e^{2d}\partial_{\cal M}({\cal E}^{\cal M}{}_{\cal A}e^{-2d})  = - {\cal F}_{\cal A}\, .
\ee

The action principle of the $O(D,D+K)$ invariant DFT, in the generalized metric formulation, is given by 
\bea
S = \int d^{2D+K}X e^{-2 d} {\cal L} \, ,
\eea
where,
\bea
{\cal L} & = &  \frac18 {\cal H}^{\cal M  N} \partial_{\cal M}{{\cal H}^{\cal K  L}} \partial_{\cal N}{{\cal H}_{\cal K  L}} - \frac12 {\cal H}^{\cal M  N} \partial_{\cal N}{{\cal H}^{\cal K  L}} \partial_{\cal L}{ {\cal H}_{\cal M  K}} \nn \\
&& + 4 {\cal H}^{\cal M  N} \partial_{\cal M}{d} \partial_{\cal N}{ d} - 2 \partial_{\cal M}{ {\cal H}^{ \cal M  N}} \partial_{\cal N}{d} - \frac12 f^{\cal M}{}_{\cal N K} {\cal H}^{\cal N P} {\cal H}^{\cal K Q} \partial_{\cal P} {\cal H}_{\cal M Q} \nn \\
&& - \frac{1}{12} f^{\cal M}{}_{\cal K P} f^{\cal N}{}_{\cal L Q} {\cal H}_{\cal M N} {\cal H}^{\cal K L} {\cal H}^{\cal P Q} \nn \\ && - \frac14 f^{\cal M}{}_{\cal N K} f^{\cal N}{}_{\cal M L} {\cal H}^{\cal K L} - \frac16 f^{\cal M N K} f_{\cal M N K} \, .
\label{MetricLag}
\eea

The Lagrangian (\ref{MetricLag}) can be written in terms of generalized fluxes \cite{Exploring} as \footnote{We are neglecting a total derivative.}
\bea
2{\cal E}_{\underline{\cal A}}{\cal F}^{\underline{\cal A}} + {\cal F}_{\underline{\cal A}}{\cal F}^{\underline{\cal A}} - \frac16 {\cal F}_{\underline{\cal ABC}} {\cal F}^{\underline{\cal ABC}} - \frac12{\cal F}_{\ov{\cal A}\underline{\cal BC}}{\cal F}^{\ov{\cal A}\underline{\cal BC}} \, .
\eea

The DFT equations of motion are given by the generalized Ricci scalar,
\bea
{\cal R} = && \frac18 {\cal H}^{\cal M  N} \partial_{\cal M}{\cal H}^{\cal K  L}\partial_{\cal N}{\cal H}_{\cal K  L} - \frac12 {\cal H}^{\cal M  N}\partial_{ \cal N}{\cal H}^{\cal K  L}\partial_{\cal L}{\cal H}_{\cal M  K} + 4 {\cal H}^{\cal M  N} \partial_{ \cal M}\partial_{\cal N}  d \nn \\ && + 4 \partial_{\cal M}{\cal H}^{\cal M  N} \partial_{\cal N} d - 4 {\cal H}^{\cal M  N} \partial_{\cal M} d \partial_{\cal N} d -  \partial_{\cal M} \partial_{\cal N} {\cal H}^{\cal M  N} - \frac12 f^{\cal M}{}_{\cal N K} {\cal H}^{\cal N P} {\cal H}^{\cal K Q} \partial_{\cal P} {\cal H}_{\cal M Q} \nn \\
&& - \frac{1}{12} f^{\cal M}{}_{\cal K P} f^{\cal N}{}_{\cal L Q} {\cal H}_{\cal M N} {\cal H}^{\cal K L} {\cal H}^{\cal P Q} - \frac14 f^{\cal M}{}_{\cal N K} f^{\cal N}{}_{\cal M L} {\cal H}^{\cal K L} \nn \\ && - \frac16 f^{\cal M N K} f_{\cal M N K} \, = 0 \, , 
\eea
and a generalized Ricci tensor
\bea
 {\cal R}_{\cal M  N} = P_{\cal M}{}^{\cal P} {\cal K}_{\cal P  Q} {\ov P}^{\cal Q}{}_{\cal N} +  {\ov P}_{\cal M}{}^{\cal P} {\cal K}_{\cal P  Q} P^{ \cal Q}{}_{\cal N} = 0 \, ,
\eea
where
\bea
 {\cal K}_{\cal M  N} & = & \frac{1}{8} \partial_{\cal M} {\cal H}^{\cal K  L} \partial_{\cal N} {\cal H}_{\cal K  L} - \frac14 \left(\partial_{\cal L} - 2 \partial_{\cal L}  d\right)\left({\cal H}^{\cal L  K} \partial_{\cal K}{\cal H}_{\cal M  N}\right) + 2 \partial_{\cal M}\partial_{\cal N}  d \nn \\ && - \frac12 \partial_{(\cal M} {\cal H}^{\cal K  L} \partial_{\cal L} {\cal H}_{\cal N)  K} + \frac12 \left(\partial_{\cal L} - 2 \partial_{\cal L}  d\right) \left( {\cal H}^{\cal K  L} \partial_{\cal( M} {\cal H}_{\cal N)  K} + {\cal H}^{\cal K}{}_{\cal( M} \partial_{\cal K} {\cal H}^{ \cal L}{}_{\cal N)}\right) \nn \\ 
 && + \frac12 f_{(\cal M L}{}^{\cal K} {\cal H}^{\cal L P} \partial_{\cal P} {\cal H}_{\cal N)K} - \frac12 f_{\cal (M L}{}^{\cal K} {\cal H}^{\cal L P} \partial_{\cal N)} {\cal H}_{\cal P K} - \frac14 f_{\cal M L}{}^{\cal K} f_{\cal NK}{}^{\cal L} \nn \\ &&
 + \frac12 e^{2d} \partial_{\cal P}(e^{-2d} {\cal H}^{\cal L P} {\cal H}_{\cal Q(M}) f_{\cal N)L}{}^{\cal Q} - \frac14 f_{\cal MK}{}^{\cal P} f_{\cal NL}{}^{\cal Q} {\cal H}^{\cal K L} {\cal H}_{\cal P Q} 
 \, .
\eea

So far we have reviewed the standard construction of heterotic DFT. As we mention before in this formulation there are no signs of a gauge field since it is embedded in the generalized metric/frame, while the gauge symmetry is embedded in the generalized diffeomorphisms.

\subsection{Heterotic Kerr-Schild ansatz in metric formulation}
The formulation of the GKSA for heterotic DFT consists of an exact and linear perturbation of a generalized background metric \cite{HetKL} 
\bea
{\cal H}_{\cal MN} = {\cal H}_{o\cal MN} + \kappa (\ov{\cal K}_{\cal M} {\cal K}_{\cal N} + {\cal K}_{\cal M} \ov{\cal K}_{\cal N} ) \, ,
\label{DFTKS}
\eea
where $\ov{\cal K}_{\cal M}= \ov{\cal P}_{\cal M}{}^{\cal N} \ov{\cal K}_{\cal N}$ and ${\cal K}_{\cal M}= {\cal P}_{\cal M}{}^{\cal N} {\cal K}_{\cal N}$ are a pair of generalized null vectors 
\bea
\eta^{\cal MN} \ov{\cal K}_{\cal M} \ov{\cal K}_{\cal N} & = & 0 \, , \nn \\ \eta^{\cal MN} {\cal K}_{\cal M} {\cal K}_{\cal N} & = & 0 \, .
\label{nulldft}
\eea 
According to (\ref{DFTKS}), the DFT projectors are perturbed as follows,
\bea
{\cal P}_{\cal MN} &=& {\cal P}_{o\cal MN} - \frac12 \kappa (\ov{K}_{\cal M} {\cal K}_{\cal N} + {\cal K}_{\cal M} \ov{\cal K}_{\cal N}) \nn \\ 
\ov{\cal P}_{\cal MN} &=& \ov{\cal P}_{\cal MN} + \frac12 \kappa (\ov{\cal K}_{\cal M} {\cal K}_{\cal N} + {\cal K}_{\cal M} \ov{\cal K}_{\cal N}) \, .
\eea
The generalized background dilaton can be perturbed with a generic $\kappa$ expansion,
\be
d = d_{o} + \kappa f\, , \qquad f = \sum_{n=0}^{\infty}\kappa^{n}f_{n} \, .
\ee

Mimicking the ordinary Kerr-Schild ansatz \cite{KS}, the generalized vectors ${\cal K}_{\cal M}$, $\ov{\cal K}_{\cal M}$ and $f$ obey some conditions in order to produce finite deformations in the DFT action and EOM's. Following the original construction of the GKSA \cite{KL} we impose,
\bea
{\ov {\cal K}}^{\cal P} \partial_{\cal P} {\cal K}^{\cal M} + {\cal K}_{\cal P} \partial^{\cal M}{\ov K}^{\cal P} - {\cal K}^{\cal P} \partial_{\cal P}{\ov {\cal K}}^{\cal M} & = & 0 \, , \nn \\ {\cal K}^{\cal P} \partial _{\cal P}  {\ov {\cal K}}^{\cal M} + \ov{\cal K}_{\cal P} \partial^{\cal M}{\cal K}^{\cal P} - \ov{\cal K}^{\cal P} \partial_{\cal P}{\cal K}^{\cal M} & = & 0 \, , 
\label{geodesic1}
\eea
and
\bea
{\cal K}^{\cal M} \partial_{\cal M}f = \ov{\cal K}^{\cal M} \partial_{\cal M}f = 0 \, .
\label{geodesic2}
\eea
Using (\ref{torsion}), we can change $\partial \rightarrow \nabla$ in (\ref{geodesic1}) obtaining, 
\bea
\ov{\cal K}^{\cal P} \nabla_{\cal P} {\cal K}^{\cal M} + {\cal K}_{\cal P} \nabla^{\cal M}\ov{\cal K}^{\cal P} - {\cal K}^{\cal P} \nabla_{\cal P}\ov{\cal K}^{\cal M} & = & 0 \, , \nn \\ {\cal K}^{\cal P} \nabla_{\cal P} \ov{\cal K}^{\cal M} + \ov{\cal K}_{\cal P} \nabla^{\cal M}{\cal K}^{\cal P} - \ov{\cal K}^{\cal P} \nabla_{\cal P}{\cal K}^{\cal M} & = & 0 \, . 
\label{geodesic11}
\eea
In order to make contact with the ordinary heterotic supergravity we need to parametrize our background fields and their perturbations. The parametrization of the generalized metric is given by
\bea
{\cal H}_{\cal M N} = \left(\begin{matrix}  g_{o}^{\mu \nu} & -  g_{o}^{\mu \rho} C_{o\rho \nu} & -  g_{o}^{\mu \rho} A_{o \rho i} \\
-  g_{o}^{\nu \rho} C_{o\rho \mu} &  g_{o\mu \nu} + C_{o\rho \mu} C_{o\sigma \nu}  g_{o}^{\rho \sigma} + A_{o\mu}{}^i \kappa_{ij} A_{o\nu}{}^j &
C_{o\rho \mu}  g_{o}^{\rho \sigma} A_{o\sigma i} + A_{o\mu}{}^j \kappa_{ji} \\
-  g_{o}^{\nu \rho} A_{o\rho i} & C_{o\rho \nu}  g_{o}^{\rho \sigma} A_{o\sigma i} +  A_{o\nu}{}^j \kappa_{ij} & \kappa_{ij} + A_{o\rho i}  g_{o}^{\rho \sigma} A_{o\sigma j}\end{matrix}\right) \ ,
\label{Gmetric}
\eea
where $C_{o\mu \nu}=b_{o\mu \nu} + \frac12 A_{o\mu}{}^{i} A_{o\nu i}$ and $\mu=0,\dots,D-1$, $i=1,\dots,K$, while the parametrization of the generalized background dilaton is given by
\bea
e^{-2d_{o}} = \sqrt{-g_{o}} e^{-2\phi_{o}}.
\eea
The generalized vectors ${\cal K}_{\cal M}$ and $\ov{\cal K}_{\cal M}$ can be parametrized as
\bea
K_{ M} = \, \frac{1}{\sqrt{2}} \left( \begin{matrix} l^{\mu} \\ - l_{\mu} - C_{o \rho \mu} l^{\rho } \\ - A_{o \rho i} {l}^{\rho} \end{matrix} \right) \, , \quad
\bar{K}_{ M} = \, \frac{1}{\sqrt{2}} \left( \begin{matrix} {\bar l}^{\mu} \\  {\bar l}_{\mu} - C_{o \rho \mu} {\bar l}^{\rho} - \sqrt{2} A_{o\mu i} j^{i} \\- A_{o\rho i} {\bar l}^{\rho} + \sqrt{2} j_{i} \end{matrix} \right) \, ,
\eea
where $l$ and $\ov l$ satisfy,
\bea
l_{\mu} l^{\mu} & = & 0 \, , \\
\ov l_{\mu} \ov l^{\mu} + j^{i} j_{i} & = & 0 \, .
\eea
These objects also satisfy geodesic conditions, which are inherited from generalized conditions at the DFT level. 

Finally, the heterotic supergravity field content is perturbed in the following way,
\bea
\label{paramsugra}
g_{\mu \nu} & = & g_{o \mu \nu} - \frac{\kappa}{1+\frac12 \kappa l.\bar{l}} l_{(\mu} \bar{l}_{\nu)} \\ b_{\mu \nu} & = & b_{o \mu \nu} - \frac{\kappa}{1+\frac12 \kappa l.\bar{l}} l_{[\mu} (\bar{l}_{\nu]} - \frac{1}{\sqrt{2}} j_{i} A_{\nu]}{}^{i} ) \\ 
A_{\mu i} & = & A_{o \mu i} + \frac{1}{\sqrt{2}} \frac{\kappa}{1+\frac12 \kappa l.\bar{l}} l_{\mu} j_{i}
\\
\phi & = & \phi_{o} + \kappa f \, ,
\label{paramsugra2}
\eea
where we keep the same notation for the perturbation of the standard dilaton. While the ordinary Kerr-Schild ansatz is based on linear perturbations on the metric tensor, the generalized Kerr-Schild ansatz contains a tower of perturbations due to $l . \ov{l} \neq 0$. Moreover, only $l$ is a null vector when the gauge sector is taking into account.

In the next part of the work we discuss about an obstruction when the generalized frame formulation is used. Particularly, we show that the background gauge field $A_{o \mu i}$ cannot be perturbed when one considers heterotic DFT written in terms of a fundamental $O(D,D+K)$ covariant frame.

\subsection{Heterotic Kerr-Schild ansatz in frame formulation}

The generalized Kerr-Schild ansatz for the DFT frame is given by,
\bea
{\cal E}_{\cal M}{}^{\ov{\cal A}} = {\cal E}_{o\cal M}{}^{\ov{\cal A}} + \frac12 \kappa {\cal E}_{o\cal N}{}^{\ov{\cal A}} {\cal K}_{\cal M} \ov{\cal K}^{\cal N} \, , \nn \\ {\cal E}_{\cal M}{}^{\underline{\cal A}} = {\cal E}_{o M}{}^{\underline {\cal A}} - \frac12 \kappa {\cal E}_{o \cal N}{}^{\underline{\cal A}} \ov{\cal K}_{\cal M} {\cal K}^{\cal N}\, .
\eea

In the frame formulation of DFT, conditions (\ref{geodesic1}) and (\ref{geodesic2}) become \cite{EA2}
\bea
{\cal K}^{\underline {\cal A}} {\cal E}_{\underline {\cal A}} \ov{\cal K}^{\overline{\cal C}} + {\cal K}^{\un {\cal A}} \ov{\cal K}^{\ov {\cal B}} {\cal F}_{\un{\cal A} \ov{\cal B}}{}^{\ov{\cal C}} & = & 0 \, , \nn \\
\ov{\cal K}^{\overline {\cal A}} {\cal E}_{\overline {\cal A}} {\cal K}^{\underline{\cal C}} + \ov{\cal K}^{\ov{\cal A}}  {\cal K}^{\un{\cal B}} {\cal F}_{\ov{\cal A} \un{\cal B}}{}^{\un{\cal C}} & = & 0 \, ,
\label{flatgeo1}
\eea
and
\bea
{\cal K}^{\underline {\cal A}} {\cal E}_{\underline {\cal A}} f = \ov{\cal K}^{\cal A} {\cal E}_{\overline{\cal A}} f = 0 \, , 
\label{flatgeo2}
\eea
where ${\cal K}_{\un{\cal A}} = {\cal E}^{\cal M}{}_{\underline{\cal A}} {\cal K}_{\cal M}$ and $\ov{\cal K}_{\overline{\cal A}} = {\cal E}^{\cal M}{}_{\overline{\cal A}} \ov{\cal K}_{\cal M}$.

Using these identifications, conditions (\ref{flatgeo1}) and (\ref{flatgeo2}) can be written as 
\bea
{\cal K}_{\underline {\cal A}} {\cal D}^{\underline {\cal A}} \ov{\cal K}^{\overline{\cal B}}  & = &  \ov{\cal K}_{\overline{\cal A}} {\cal D}^{\overline{\cal A}} {\cal K}^{\underline{\cal B}} =  0 \, , \nn \\ {\cal K}_{\underline{\cal A}} {\cal D}^{\underline{\cal A}} f & = & {\cal K}_{\overline{\cal A}} {\cal D}^{\overline{\cal A}}f = 0 \, ,
\eea
where ${\cal D}_{\cal A}$ is a background Lorentz covariant derivative.

The parametrization of the generalized background frame is given by
\bea
{\cal E}^{\cal M}{}_{\cal A} = \frac{1}{\sqrt{2}}\left(\begin{matrix}-{ e}_{o\mu a}-C_{o\rho\mu} { e}_{o}^{\rho }{}_{a} &  { e}_{o}^{\mu }{}_{a} & -A_{o\rho}{}^i { e}_{o}^{\rho }{}_{{a}}\, , \\ 
{\overline e}_{o \mu a}-C_{o\rho \mu}{} {\overline e}_{o}^{\rho }{}_{{a}}& {\overline e}_{o}^\mu{}_{a}&-A_{o\rho}{}^i  {\overline e}_{o}^\rho{}_{a} \\
\sqrt{2} A_{o\mu i}e^i{}_{\overline i} &0&\sqrt{2} e^i{}_{\overline i} \end{matrix}\right)  \, ,
\label{HKparam}
\eea
and we impose the standard gauge fixing for the double Lorentz group,
\bea
e_{o\mu a} \eta^{ab} e_{o\nu b} = {\overline e}_{o\mu a} \eta^{ab} {\overline e}_{o\nu b} = {g}_{o \mu \nu} \, , 
\label{gf}
\eea
with $\eta_{ab}$ the ten dimensional flat metric, $a,b=0,\dots, D-1$. 

Since ${\cal E}^{\mu}{}_{\bar i}$ cannot be perturbed, we are forced to impose $j_{i}=0$ from the very beginning. Then, the parametrization of the generalized vector fields is
\bea
{\cal K}_{\cal M} = \, \frac{1}{\sqrt{2}} \left( \begin{matrix} l^{\mu} \\ - l_{\mu} - C_{o \rho \mu} l^{\rho } \\ - A_{oi \rho} {l}^{\rho} \end{matrix} \right) \, , \quad
\ov{\cal K}_{\cal M} = \, \frac{1}{\sqrt{2}} \left( \begin{matrix} {\bar l}^{\mu} \\  {\bar l}_{\mu} - C_{o \rho \mu} {\bar l}^{\rho} \\- A_{oi \rho} {\bar l}^{\rho} \end{matrix} \right) \,  
\eea
In this framework, both $l$ and $\ov{l}$ are null vectors in order to preserve the condition
\bea
{\cal E}^{\mu}{}_{\bar i} = 0 \, .
\eea
Then, it is possible to recover perturbations for the supergravity fields but part of the perturbations of the b-field and the full perturbation for the gauge field is missed.

In the next section we rewrite heterotic DFT in terms of $O(D,D)$ fundamental fields in order to present a double Yang-Mills formulation of this theory. The leading order terms of our construction is closely related to the construction given in \cite{PK}.  
 
\section{Double Yang-Mills formulation}
\label{DYM}
We start by considering a splitting of heterotic DFT. We parametrize the $O(D,D+K)$ generalized frame in terms of $O(D,D)$ multiplets, i.e.,
\be
{\cal E}_{\cal M}{}^{\cal A}  = 
\left(\begin{matrix}{\cal E}_{M}{}^{A}& {\cal E}_{M}{}^{\bar \alpha}\\ 
{\cal E}_\alpha{}^A&{\cal E}_{\alpha}{}^{\bar \alpha} 
\end{matrix}\right)
= \left(\begin{matrix}({\chi}^{\frac12})_{M}{}^{N} E_{N}{}^{A}& -A_{M}{}^{\beta} e_{\beta}{}^{\ov \alpha}\\ 
{A}^{M}{}_\alpha E_{M}{}^{A}&({\Box}^{\frac12})_{\alpha}{}^{\beta} e_{\beta}{}^{\ov \alpha} 
\end{matrix}\right) \ , \label{Extframe}
\ee
where $e_{\alpha}{}^{\ov \a}$ satisfies
\bea
e_{\alpha}{}^{\ov \a} \kappa_{\ov \a \ov \b} e_{\b}{}^{\ov \b} & = & \kappa_{\a \b} \, , \\
e^{\alpha}{}_{\ov \a} \kappa_{\a \b} e^{\b}{}_{\ov \b} & = & \kappa_{\ov \a \ov \b}
\, .
\eea
The matrices $\chi_{MN}$ and $\Box_{\a \b}$ are defined in the following way, 
\bea
\label{chiex}
{\chi}_{M N}=\eta_{M N} - A_{M}{}^{\alpha} A_{N \alpha}  \, , \quad
{\Box}_{\a \b}=\kappa_{\a \b} - A_{M \alpha} A^{M}{}_{\beta}  \, ,
\eea
and we also impose
\bea
E^{M \ov A} A_{M \a} = 0 \, 
\eea
which requires a gauge fixing. The projectors ${\cal P}_{\cal M N}$, $\overline{\cal P}_{\cal M N}$ and ${\cal P}_{\cal A B}$, $\overline{\cal P}_{\cal A B}$ are decomposed as 
\be
{\cal P}_{\cal M N}  =\left(\begin{matrix}{P}_{M N}&  0 & \ \ 0 \\ 
0 & 0 & \ \ 0 \\
0 & 0 & \ \ 0 
\end{matrix}\right) \, , \quad \ov {\cal P}_{\cal M N}  =\left(\begin{matrix}0 &  0 & 0\\ 
0 & {\ov P}_{M N} & 0 \\
0 & 0 & {\kappa}_{\a \b} 
\end{matrix}\right) \, ,
\ee
\be
{\cal P}_{\cal A B}  =\left(\begin{matrix}{P}_{A B}&  0 & \ \ 0 \\ 
0 & 0 & \ \ 0 \\
0 & 0 & \ \ 0 
\end{matrix}\right) \, , \quad \ov {\cal P}_{\cal A B}  =\left(\begin{matrix}0 &  0 & 0\\ 
0 & {\ov P}_{A B} & 0 \\
0 & 0 & {\kappa}_{\ov \a \ov \b} 
\end{matrix}\right) \, .
\ee

In \cite{gbdr}, the ansatz (\ref{Extframe}) was used to obtain $\alpha'$-corrections. There, the authors considered that the $A_{M \a}$ is not a fundamental field and, consequently, they identified it with some projections of the DFT fluxes. Here we follow a different philosophy. We consider $A_{M \a}$ as a fundamental field and we construct a double Yang-Mills action. Moreover, the parametrization of this field gives raise to the ordinary Yang-Mills connection $A_{\mu i}$. As we previously mentioned, $\Big\{ E_{M A}, A_{M \alpha}, d \Big\}$ are now the fundamental fields of this alternative formulation of heterotic DFT. Similarly, the symmetry rules given by $\hat \xi_{\cal M}$ and $\Gamma_{\cal A \cal B}$ must be decomposed in a consistent way.   

The symmetry transformations of the $O(D,D+K)$ fields are given by
\bea
\delta {\cal E}_{\cal M A} & = &  \hat \xi^{\cal N} \partial_{\cal N} {\cal E}_{\cal M A} + (\partial_{\cal M} \hat \xi^{\cal P} - \partial^{\cal P} \hat \xi_{\cal M}) {\cal E}_{\cal P A} + f_{\cal M N P} \hat \xi^{\cal N} {\cal E}^{\cal P}{}_{\cal A} + {\cal E}_{\cal M B} \Gamma^{\cal B}{}_{\cal A}  \, , \\
\delta d & = & \hat \xi^{\cal N} \partial_{\cal N} d - \frac{1}{2} \partial_{\cal M} \hat \xi^{\cal M} \, ,
\eea
where $f_{\cal MNP}$ only takes values when ${\cal MNP}=\alpha \beta \gamma$. The $\partial_{\cal M}$ derivative  is split according to $\partial_{\cal M}=(\partial_{M},0)$ and the $O(D,D)$ strong constraint is
\bea
\partial_{M} \blacktriangledown \partial^{M} \blacktriangle & = & 0 \, , \\
\partial_{M}(\partial^{M} \blacktriangledown) & = & 0 \, .
\eea
The generalized diffeomorphisms parameter is split as 
\bea
\hat \xi^{\cal M} & = & (\hat{\xi}^{M}, \lambda^{\alpha}) \, ,
\eea
with $\lambda_{\alpha}$ a gauge parameter. 

Some of the components of the double Lorentz parameter, ${\Gamma_{\cal AB}}$, require a gauge fixing to ensure $\delta {\cal E}_{\alpha}{}^{\ov A} = 0$ and $\delta e_{i}{}^{\ov i}=0$. From the former we find,
\bea
\Gamma_{\ov \b}{}^{\ov A} = (\Box^{-1/2})^{\alpha}{}_{\beta} e^{\beta}{}_{\ov \b} (\partial^{P} \lambda_{\alpha}) E_{P}{}^{\ov A} \, . 
\label{gfnew}
\eea
Similarly, we can now demand $\delta {\cal E}_{\alpha}{}^{\ov \alpha} = \delta(\Box^{\frac12})_{\alpha}{}^{\beta} e_{\beta}{}^{\ov \alpha}$, and then $\delta e_{\a}{}^{\ov \a}=0$. The previous condition is satisfied if
\bea
\Gamma_{\ov \b \ov \a} = (\Box^{-1/2})^{\a}{}_{\b} e^{\b}{}_{[\ov \b} \Big(- \hat \xi_{P} \partial^{P}{\cal E}_{\alpha \ov \alpha]} + \partial^{P} \lambda_{\alpha} {\cal E}_{P \ov \a]}  - f_{\alpha \delta \gamma} \lambda^{\delta} {\cal E}^{\gamma}{}_{\ov \a]} \Big) \, .
\label{gfpar}
\eea
Equations (\ref{gfnew}) and (\ref{gfpar}) are the gauge fixing conditions that we need to write heterotic DFT in terms of fields which are in representations of $O(D,D)$. In the next part of the work we explore the transformation rules for these fields. 

\subsection{Transformation rules}
We start by defining the following field
\bea
C_{M \alpha} = - A_{M}{}^{\beta} (\Box^{-\frac12})_{\beta \a}
\eea 
which is constrained by $E_{M}{}^{\ov A} C_{M \alpha}=0$. We also define
\bea
\Delta_{\alpha \beta} & = & \kappa_{\a \b} + C_{M \alpha} C^{M}{}_{\b} \\
\Theta_{MN} & = & \eta_{MN} + C_{M}{}^{\a} C_{N \a} \, .
\eea
The previous objects satisfy
\bea
\Delta_{\alpha \beta} = (\Box^{-1}){}_{\alpha \beta} \ , \ \ \ \ \Theta_{M N} = (\chi^{-1}){}_{M N} \ .
\label{relAC}
\eea
There relations are useful to write the action principle in terms of $A_{M \alpha}$ or in terms of $C_{M \alpha}$.

Now we inspect the transformation of $\delta {\cal E}_{M \ov A}=\delta {E}_{M \ov A}$. In terms of the $C_{M \alpha}$ field we have,
\bea
\delta {E}_{M \ov A} = {\cal L}_{\hat \xi} E_{M \ov A} + E_{M}{}^{\ov B} \Lambda_{\ov B \ov A} + C_{M}{}^{\gamma} \partial^{P}\lambda_{\gamma} E_{P \ov A} \, ,
\eea
where we have identified
\bea
\Gamma_{\ov A \ov B} = \Lambda_{\ov A \ov B} \, .
\eea
At this point we need to inspect the other generalized frame projection $\delta E_{M \un A}$, 
\bea
\delta {\cal E}_{M \un A} = \delta(\chi^{\frac12})_{M}{}^{N} E_{N}{}^{\un A} + (\chi^{\frac12})_{M}{}^{N} \delta E_{N}{}^{\un A} \, .
\eea
From the previous expression we find,
\bea
\delta E_{N \un A} = {\cal L}_{\hat \xi} E_{N \un A} + E_{N}{}^{\un B} \Gamma_{\un B \un A} - (\Theta^{\frac12})^{M}{}_{N} \Big(\partial_{M} \lambda^{\a} C^{R}{}_{\beta} (\Delta^{\frac12})^{\beta}{}_{\a} + \delta_{\Lambda} (\Theta^{-\frac12})_{M}{}^{R} \Big) E_{R \un A} \, ,
\label{trframedown}
\eea
which need a parameter redefinition since we want (\ref{trframedown}) to have a generalized Green-Schwarz form. Then, we must define
\bea
Q_{N}{}^{R} = (\Theta^{\frac12})^{M}{}_{N} \Big( \partial_{M} \lambda^{\a} C^{R}{}_{\beta} (\Delta^{\frac12})^{\beta}{}_{\a} +  \delta_{\Lambda} (\Theta^{-\frac12})_{M}{}^{R} \Big) E_{R \un A} \, ,
\eea
and $S_{MN}=\ov P_{M}{}^{P} \partial_{P} \lambda^{\alpha} C_{N \alpha}-Q_{MN}$, in order to identify
\bea
\Gamma_{\un A \un B} = \Lambda_{\un A \un B} - E^{M}{}_{\un A} S_{MN} E^{N}{}_{\un B} \, .
\eea
In terms of the previous parameter, the transformation of this component takes its expected form,
\bea
\delta E_{N \un A} = {\cal L}_{\hat \xi} E_{N \un A} + E_{N}{}^{\un B} \Lambda_{\un B \un A} - \partial_{\ov N} \lambda^{\alpha} C^{R}{}_{\alpha} E_{R \un A} \, .
\label{frameun}
\eea
Finally, the transformation of $C_{M \alpha}$ reads
\bea
\delta C_{M}{}^{\gamma} = {\cal L}_{\hat \xi} C_{M}{}^{\gamma} + \partial_{M} \lambda^{\gamma} - (\Box^{-1})^{\gamma \alpha} \partial_{\ov M} \lambda_{\alpha} + C_{M}{}^{\alpha} \partial^{P}\lambda_{\alpha} C_{P}{}^{\gamma} - C_{M}{}^{\alpha} f_{\alpha \beta}{}^{\gamma} \lambda^{\beta} \, . 
\label{Ctransf}
\eea
So far we have computed the transformation rules for all the fundamental fields of this double Yang-Mills formulation of heterotic DFT. 

\subsection{Flux formulation}
The $O(D,D+K)$ invariant fluxes \cite{Exploring} can be easily decomposed in terms of $O(D,D)$ multiplets using (\ref{Extframe}),
\bea
{\cal F}_{\un A \un B \un C} & = & 3 \sqrt{2} (\chi^{\frac12})^{M}{}_{P} E^P{}_{[\un A} \Big( \partial_{M}((\chi^{\frac12})^{N Q}E_{Q \un B})(\chi^{\frac12})_{NR} E^{R}{}_{\un C]} + \partial_{M} A_{\un B}{}^{\a} A_{\un C] \a} \Big) \nn \\
&& + \sqrt{2} f^{\a \b \gamma} A_{\un A \a} A_{\un B \b} A_{\un C \gamma}  \, ,
\eea
\bea
{\cal F}_{\ov A \un B \un C} & = & \sqrt{2} E^M{}_{\ov A} \partial_{M}((\chi^{\frac12})^{NQ} E_{Q\un B})(\chi^{\frac12})_{NR} E^{R}{}_{\un C} + 2 \sqrt{2} (\chi^{\frac12})^{M}{}_{P} E^P{}_{[\un C} \partial_{M}(E^{N}{}_{\ov A})(\chi^{\frac12})_{NR} E^{R}{}_{\un B]} \nn \\ && + 2 \sqrt{2} E^M{}_{\ov A} \partial_{M}(A_{[\un B}{}^{\a})A_{\un C] \a} \, ,
\eea
\bea
{\cal F}_{\ov \a \un B \un C} & = & - \sqrt{2} A^M{}_{[\a} e^{\a}{}_{\ov \a} \partial_{M}((\chi^{\frac12})^{NQ} E_{Q\un B})(\chi^{\frac12})_{NR} E^{R}{}_{\un C} + 2 \sqrt{2} (\chi^{\frac12})^{M}{}_{P} E^P{}_{[\un C} \partial_{M}((\Box^{\frac12})_{\a \b} e^{\b}{}_{\ov \a}) A_{\un B]}{}^{\a} \nn \\ && + \sqrt{2} A^M{}_{\b} e^{\b}{}_{\ov \a} \partial_{M} A_{\un B}{}^{\a} A_{\un C \a} + \sqrt{2} f^{\a \b \gamma} (\Box^{\frac12})^{\a \delta} e_{\delta \ov \a} A_{\un B \b} A_{\un C \gamma} \, ,  
\eea
\bea
{\cal F}_{\un A} & = & \sqrt{2} \partial_{M}((\chi^{\frac12})^{MP} E_{P \un A}) - 2 \sqrt{2} (\chi^{\frac12})^{MP} E_{P \un A} \partial_{M}d \, .
\eea
where $A_{\un B \alpha}=A_{M \alpha} E^{M}{}_{\un B}$. Other projections can be decomposed in the same way. Here we prioritize the ones which appear in the heterotic DFT action.

\subsection{Action principle}

The components of the $O(D,D+K)$ generalized metric can be written in terms of the following $O(D,D)$ multiples $\left\{H_{M N},A_{M \alpha} \right\}$ similarly to the decomposition given in \cite{HSZheter}. According to our conventions,
\bea
{\cal H}_{M N} & = & H_{M N} + 2 A_{M \a} A_{N}{}^{\a} = H_{M N} + 2 \eta_{M N} - 2 (\Theta^{-1})_{M N} \, , \\
{\cal H}_{M \beta} & = & - 2 (\chi^{\frac12})_{M}{}^{P} A_{P \b} = 2 C_{M \beta} \, , \\
{\cal H}_{\alpha N} & = & - 2 (\chi^{\frac12})_{N}{}^{Q} A_{Q \a} = 2 C_{N \a} \, , \\
{\cal H}_{\a \b} & = & \kappa_{\a \b} - 2 A^{M}{}_{\a} A_{M \b} = - \kappa_{\a \b} + 2 (\Delta^{-1})_{\a \b} \, .
\eea
The decomposition of the DFT Lagrangian is given by
\begin{eqnarray}
&\mathcal{L} =& \frac{1}{8} \mathcal{H}^{MN} \partial_M \mathcal{H}^{KL}  \partial_N \mathcal{H}_{KL} + \frac{1}{4} \mathcal{H}^{MN} \partial_M \mathcal{H}^{K\alpha}  \partial_N \mathcal{H}_{K\alpha} + \frac{1}{8} \mathcal{H}^{MN} \partial_M \mathcal{H}^{\alpha\beta}  \partial_N \mathcal{H}_{\alpha\beta} \\
&& -  \frac{1}{2} \mathcal{H}^{MN} \partial_N \mathcal{H}^{KL}  \partial_L \mathcal{H}_{MK} - \frac12  \mathcal{H}^{MN} \partial_N \mathcal{H}^{\beta K}  \partial_K \mathcal{H}_{M\beta} - \frac{1}{2} \mathcal{H}^{\alpha K} \partial_K \mathcal{H}^{MN}  \partial_N \mathcal{H}_{\alpha M} \nonumber\\
&&- \frac{1}{2} \mathcal{H}^{\alpha K} \partial_K \mathcal{H}^{\beta N}  \partial_N \mathcal{H}_{\alpha \beta} + 4 \mathcal{H}^{MN} \partial_M d \partial_N d - 2 \partial_M \mathcal{H}^{MN} \partial_N d - \frac{1}{2} f_{\beta\gamma}^{\alpha} \mathcal{H}^{\beta M} \mathcal{H}^{\gamma N} \partial_M \mathcal{H}_{\alpha N} \nonumber\\
&& -  \frac{1}{2} f_{\beta\gamma}^{\alpha} \mathcal{H}^{\beta M} \mathcal{H}^{\gamma \lambda} \partial_M \mathcal{H}_{\alpha \lambda} - \frac{1}{12} f_{\beta\gamma}^{\alpha} f_{\mu\rho}^{\lambda} \mathcal{H}_{\alpha\lambda} \mathcal{H}^{\beta\mu} \mathcal{H}^{\gamma\rho}  - \frac{1}{4} f_{\beta\gamma}^{\alpha} f_{\alpha\lambda}^{\beta} \mathcal{H}^{\gamma\lambda} - \frac{1}{6} f^{\alpha\beta\gamma} f_{\alpha\beta\gamma} \ , \nn
\label{Lagrangian}
\end{eqnarray}
and the double Yang-Mills action principle takes the following form,
\bea
S & = & \int d^{2D}X e^{-2d} \Big(\frac{1}{8} (H^{MN} - 2 (\Theta^{-1})^{M N}) \partial_M (H^{KL} - 2 (\Theta^{-1})^{K L})  \partial_N (H_{KL} - 2 (\Theta^{-1})_{KL}) \nn \\ && + (H^{MN} - 2 (\Theta^{-1})^{M N}) \partial_M C^{K\alpha}  \partial_N C_{K\alpha} + \frac{1}{2} (H^{MN} - 2 (\Theta^{-1})^{M N}) \partial_M (\Delta^{-1})^{\alpha\beta}  \partial_N (\Delta^{-1})_{\alpha\beta} \nonumber\\
&& -  \frac{1}{2} (H^{MN} + 2 \eta^{M N} - 2 (\Theta^{-1})^{M N}) \partial_N (H^{KL} - 2 (\Theta^{-1})^{K L})  \partial_L (H_{MK} - 2 (\Theta^{-1})_{M K}) \nn \\ && - 2  (H^{MN} + 2 \eta^{M N} - 2 (\Theta^{-1})^{M N}) \partial_N C^{K \beta}  \partial_K C_{M\beta} - 2 C^{K \alpha} \partial_K(H^{MN} - 2 (\Theta^{-1})^{M N})  \partial_N C_{M \alpha} \nonumber\\
&&- 4 C^{K \alpha} \partial_K C^{N \beta}  \partial_N (\Delta^{-1})_{\alpha \beta} + 4 (H^{MN} - 2 (\Theta^{-1})^{M N}) \partial_M d \partial_N d - 2 \partial_M(H^{MN} - 2 (\Theta^{-1})^{M N}) \partial_N d \nn \\ && - 4 f_{\beta\gamma}^{\alpha} C^{M \beta} C^{N \gamma} \partial_M C_{N \alpha} -  2 f_{\beta\gamma}^{\alpha} C^{M \beta} (-\kappa^{\gamma \lambda} + 2 (\Delta^{-1})^{\gamma \lambda}) \partial_M (\Delta^{-1})_{\alpha \lambda} \nn \\ && - \frac{1}{12} f_{\beta\gamma}^{\alpha} f_{\mu\rho}^{\lambda} (-\kappa_{\alpha\lambda} + 2 (\Delta^{-1})_{\alpha\lambda}) (-\kappa^{\beta\mu} + 2 (\Delta^{-1})^{\beta\mu}) (-\kappa^{\gamma \rho} + 2 (\Delta^{-1})^{\gamma \rho}) \nn \\ && - \frac{1}{4} f_{\beta\gamma}^{\alpha} f_{\alpha\lambda}^{\beta} (-\kappa^{\gamma\lambda} + 2 (\Delta^{-1})^{\gamma \lambda})  - \frac{1}{6} f^{\alpha\beta\gamma} f_{\alpha\beta\gamma} \Big) \ .
\label{LagC}
\eea
The dynamics can be computed considering variations on the previous Lagrangian or splitting the equations of motion from the $O(D,D+K)$ perspective. Using the latter method the equation of motion for the $O(D,D)$ dilaton reads
\bea
{\cal R}&=&0\, , \label{eomR}
\eea
or, equivalently, ${\cal L}=0$. The equation of motion for the $O(D,D)$ metric reads
\bea
&& {\cal K}_{M N} - (H_{M}{}^{P} + 2 \delta_{M}{}^{P} - 2 (\Theta^{-1})_{M}{}^{P}) {\cal K}_{Q P} (H^{Q}{}_{N} + 2 \delta^{Q}{}_{N} - 2 (\Theta^{-1})^{Q}{}_{N}) \nn \\ && - 4 C_{M}{}^{\alpha} {\cal K}_{\alpha Q} (H^{Q}{}_{N} + 2 \delta^{Q}{}_{N} - 2 (\Theta^{-1})^{Q}{}_{N}) - 4 C_{M}{}^{\alpha} {\cal K}_{\alpha \beta} C_{N}{}^{\beta}  =  0 \, , \label{eomRMN}
\eea
and the equation of motion for the gauge field is given by
\bea
&& {\cal K}_{M \alpha} - 2 (H_{M}{}^{P} + 2 \delta_{M}{}^{P} - 2 (\Theta^{-1})_{M}{}^{P}) {\cal K}_{P Q} C^{Q}{}_{\alpha} \nn \\ && - (H_{M}{}^{P} + 2 \delta_{M}{}^{P} - 2 (\Theta^{-1})_{M}{}^{P}) {\cal K}_{P \beta} (-\kappa^{\beta}{}_{\alpha} + 2 (\Delta^{-1})^{\beta}{}_{\alpha}) \nn \\ && - 2 C_{M}{}^{\gamma} {\cal K}_{\gamma \beta} (-\kappa^{\beta}{}_{\alpha} + 2 (\Delta^{-1})^{\beta}{}_{\alpha})  - 4 C_{M}{}^{\gamma} {\cal K}_{\gamma Q} C^{Q}{}_{\alpha} = 0 \, ,\label{eomRMa}
\eea
where
\bea
&\mathcal{K}_{MN} =& \frac{1}{8} \partial_M ({H}^{KL} - 2 (\Theta^{-1})^{K L}) \partial_N ({H}_{KL} - 2 (\Theta^{-1})_{K L})  + \partial_M C^{K\alpha} \partial_N C_{K\alpha}  \nonumber\\
&&  - \frac{1}{4} \big(\partial_L - 2 \partial_L d\big)  \big(({H}^{KL} - 2 (\Theta^{-1})^{K L}) \partial_K({H}_{MN} - 2 (\Theta^{-1})_{M N})\big) + 2 \partial_M \partial _N d \nonumber\\
&& - \frac{1}{2} \partial_{(M} ({H}^{KL} - 2 (\Theta^{-1})^{K L}) \partial_L ({H}_{N)K} - 2 (\Theta^{-1})_{N) K})  - 2 \partial_{(M} C^{L \alpha} \partial_L C_{N)\alpha} \nonumber\\
&& + \frac{1}{2} \big(\partial_L - 2 \partial_L d\big) \big(({H}^{KL} + 2 \eta^{K L} - 2 (\Theta^{-1})^{K L}) \partial_{(M} ({H}_{N)K} - 2 (\Theta^{-1})_{N) K}) \nn \\ && + 4 C^{L \alpha} \partial_{(M} C_{N)\alpha} + ({H}_{(MK} + 2 \eta_{(M K} - 2 (\Theta^{-1})_{(M K}) \partial^K ({H}^{L}{}_{N)} - 2 (\Theta^{-1})^L{}_{N)}) \big) \nn \\ && +  \frac{1}{2} \partial_M (\Delta^{-1})^{\alpha \beta} \partial_N (\Delta^{-1})_{\alpha \beta} \, ,
\eea
\bea
&\mathcal{K}_{M\alpha} =& - \frac{1}{2} \big(\partial_L - 2 \partial_L d\big)  \big(({H}^{LK} - 2 (\Theta^{-1})^{LK}) \partial_K C_{M\alpha}\big) - \frac12 \partial_{M} ({H}^{LK} - 2 (\Theta^{-1})^{LK}) \partial_L C_{K \alpha} \nn \\ && - \partial_{M} C^{L \beta} \partial_L (\Delta^{-1})_{\alpha \beta} + \frac{1}{2} \big(\partial_L - 2 \partial_L d\big) \big(({H}^{LK} + 2 \eta^{L K} - 2 (\Theta^{-1})^{LK}) \partial_{M} C_{K \alpha} \nn \\ && + 2 C^{L \beta} \partial_{M} (\Delta^{-1})_{\alpha \beta} + 2 ({H}_{M}{}^{K} + 2 \delta_{M}{}^{K} - 2 (\Theta^{-1})_{M}{}^{K}) \partial_K C^L{}_{\alpha )} \big) \nonumber\\
&&+ e^{2d} \partial_P \Big(e^{-2d} C^{P \beta} C_{M \gamma}\Big) f_{\alpha \beta}^\gamma \, ,
\eea
and
\bea
&\mathcal{K}_{\alpha \beta} =&  - \frac{1}{4} \big(\partial_L - 2 \partial_L d\big)  \big(({H}^{LK} - 2 (\Theta^{-1})^{LK}) \partial_K (\Delta^{-1})_{\alpha \beta}\big)  + 2 \big(\partial_L - 2 \partial_L d\big) \big(C^{K}{}_{(\alpha} \partial_K C^L{}_{\beta)} \big) \nonumber\\
&& + f_{( \alpha\gamma}^{\lambda} C^{P \gamma} \partial_P (\Delta^{-1})_{\beta ) \lambda} - \frac{1}{4} f_{\alpha\gamma}^{\lambda} f_{\beta\lambda}^{\gamma} + e^{2d} \partial_P \Big(e^{-2d} C^{P \gamma} (\Delta^{-1})_{\lambda(\alpha}\Big) f_{\beta )\gamma}^{\lambda}  \nn \\ && - \frac{1}{4} f_{\alpha\gamma}^{\lambda} f_{\beta\rho}^{\sigma} (-\kappa^{\gamma \rho} + 2 (\Delta^{-1})^{\gamma \rho}) (- \kappa_{\lambda \sigma} + 2 (\Delta^{-1})_{\lambda \sigma}) \, .
\eea
In the next subsection we explore the leading order terms of this double Yang-Mills formulation of heterotic DFT.

\subsection{Leading order terms}

This formulation of heterotic DFT contains an infinite expansion of gauge fields in the action principle. Let us focus in the leading order terms. In this limit the non-covariant transformation of the gauged vector is given by \cite{PK},
\bea
\delta_{\textrm{non-cov}} A_{M}{}^{\gamma} = - \partial_{\underline M} \lambda^{\gamma} + \mathcal{O}(A^2) \, .
\eea
A generic DFT gauged vector transforms as
\bea
\delta_{\lambda} V^{\gamma} = - A_{M}{}^{\alpha} f_{\alpha \beta}{}^{\gamma} \lambda^{\beta} \, ,
\label{leadingA}
\eea
and hence a projected covariant derivative can be defined in the following form,
\bea
\nabla_{\un M} V^{\gamma} = \partial_{\un M} V^{\gamma} - V^{\alpha} f_{\alpha \beta}{}^{\gamma} A_{\un M}{}^{\beta} \, .
\eea
Interestingly enough, the combination $\partial_{\ov M} V^{\gamma}$ cannot be put in a covariant form using the gauge connection. Moreover, when deriving a more general object one needs to consider the full DFT covariant derivative,
\bea
\nabla_{\un M} V_{N}{}^{\gamma} = \partial_{\un M} V_{N}{}^{\gamma} - \Gamma_{\un M N}{}^{P} V_{P}{}^{\gamma} - V_{N}{}^{\alpha} f_{\alpha \beta}{}^{\gamma} A_{\un M}{}^{\beta} \, ,
\eea
where $\Gamma_{MNP}$ is the generalized affine connection of DFT, which is not fully-determined \cite{RiemannDFT}.

The curvature for the generalized gauge connection is given by
\bea
F_{M N}{}^{\gamma} = 2 \partial_{[M} A_{N]}{}^{\gamma} + f^{\gamma}{}_{\alpha \beta} A_{M}{}^{\alpha} A_{N}{}^{\beta} \, .
\eea
The previous object transforms covariantly with respect to (\ref{leadingA}) upon considering generalized Jacobi identities for the structure constant, in agreement with the  constraints of the heterotic DFT formulation \cite{HetDFT}.  

The transformation rule of the generalized metric with respect to gauge transformations is given by
\bea
\delta_{\lambda} H_{M N} = 4 C_{(\un M}{}^{\a} \partial_{\ov N)} \lambda_{\a} \, = - 4 A_{(\un M}{}^{\a} \partial_{\ov N)} \lambda_{\a}+ \mathcal{O}(A^3).
\label{GSH}
\eea
Since the previous transformation is non-covariant, only $F^2$ terms considering $\eta$ contractions are fully covariant. The leading order term of (\ref{GSH}) is a generalized Green-Schwarz transformation when one considers $A_{M \a}$ as a fundamental field. Upon parametrization, the b-field inherits this kind of transformation, as well as the metric tensor, which must be redefined. We will return to this point when we parametrize the theory.

In order to construct the leading order Lagrangian, we focus on Abelian terms such that their variation are linear in the gauge field. The leading order terms of the double Yang-Mills action are given by
\bea
S & = & \int d^{2D}X e^{-2d} (\frac{1}{8} {H}^{MN} \partial_M {H}^{KL}  \partial_N {H}_{KL} -  \frac{1}{2} {H}^{MN} \partial_N {H}^{KL}  \partial_L {H}_{MK} \nn \\ && + 4 {H}^{MN} \partial_M d \partial_N d - 2 \partial_M {H}^{MN} \partial_N d + H^{MN} \eta^{KL} F_{LM}^{\alpha} F_{KN \alpha} \nn \\ && - 2 A^{N \alpha} (\partial_{N}H^{KL}) (\partial_{L}A_{K \a}) - 2 H^{M N} \partial_{N}H^{K L} (\partial_{L} A_{(K}{}^{\a}) A_{M) \a} \nn \\ && 
+ \frac14 A_{M \a} A_{N}{}^{\a} \partial^{M} H^{KL} \partial^{N}H_{K L} - A^{M}{}_{\a} A^{N \a} \partial_{N}H^{KL} \partial_{L} H_{MK} \nn \\ &&
+ H_{M N} \partial^{M}A^{K}{}_{\a} A^{L \a} \partial^{N}H_{K L} - 2 H^{M N} \partial_{N} A^{(K \a} A^{L)}{}_{\a} \partial_{L} H_{MK}
\nn \\ && + 8 A^{M \alpha} A^{N}{}_{\a} (\partial_{M}d) (\partial_{N}d) - 8 (\partial_{M}A^{(M \a}) A^{N)}{}_{\a} (\partial_{N}d) \, + \dots) \, .
 \eea
In this limit $A_{M \a}=-C_{M \a}$. Since the combination $H^{MN} \eta^{KL} F_{LM}^{\alpha} F_{KN \alpha}$ is not fully covariant, extra pieces are required to construct the invariant action. Following the same logic, the ungauged part of the Lagrangian is not gauge invariant because of the generalized Green-Schwarz mechanism. The double gauge field, $A_{M \alpha}$ transforms as a gauge connection to   leading order in gauge fields. Then, part of the leading order Lagrangian can be written in terms of its curvature but also non-covariant contributions appear. 

So far we have shown that heterotic DFT can be formulated as a deformed double Yang-Mills theory. In the next section we parametrize this theory and show the required field redefinition in order to obtain the standard heterotic supergravity. 

\subsection{Parametrization}
We parametrize the fundamental fields as follows,
\bea
{H}_{MN} &=& \left(\begin{matrix}\bar g^{\mu \nu} & - \bar g^{\mu \rho}  b_{\rho \nu} \\ b_{\mu \rho} \bar g^{\rho \nu} & \bar g_{\mu \nu} - b_{\mu \rho} \bar g^{\rho \sigma} b_{\sigma \nu} \end{matrix}\right)\ , \ \ \ \ \ \ \ \  \\
{C}_M{}^\alpha &=& \frac 1 2 \left(\begin{matrix} - \bar g^{\mu \rho} A_{\rho}{}^i \\ - b_{\mu \rho} \bar g^{\rho \sigma} A_{\sigma}{}^i + A_{\mu}{}^i\end{matrix}\right)\ .
\eea
The metric tensor $\bar g_{\mu \nu}$ inherits an anomalous gauge transformation from (\ref{GSH}) and the following field redefinition is required,
\bea
\bar g_{\mu \nu} = g_{\mu \nu} + \frac 1 2 A_{\mu}{}^i A_{\nu i}\ .
\label{redef}
\eea
This procedure is analogous to the gravitational Green-Schwarz mechanism where the metric tensor is redefined considering terms proportional to the spin connection \cite{Tduality}. 

The generalized dilaton is parametrized in the following way,
\bea
e^{-2d} & = & \sqrt{-\tilde g} e^{-2 \tilde \phi} = \sqrt{- g} e^{-2 \phi}\ ,
\eea
since the metric redefinition (\ref{redef}) implies a dilaton redefinition to obtain the standard integral measure. The transformation rules for the supergravity fields take their standard forms,
\bea
\delta g_{\mu \nu} &=& L_\xi  g_{\mu \nu}\ ,\\
\delta b_{\mu \nu} &=& L_\xi b_{\mu \nu} + 2 \partial_{[\mu} \zeta_{\nu]} -   \partial_{[\mu}\lambda^i A_{\nu]i}\ ,\\
\delta A_{\mu}{}^i &=& L_\xi A_{\mu}{}^i + \partial_{\mu} \lambda^i - f_{jk}{}^i \lambda^j A_{\mu}{}^k\ ,\\
\delta \phi &=& L_\xi \phi\ ,
\eea
where we use $L_\xi$ for the ordinary Lie derivative and $\hat \xi_{\mu}=\zeta_{\mu}$ for the abelian gauge transformation of the b-field. Finally, the action is given by
\bea
S=\int d^{D}x \sqrt{-g} e^{-2\phi} \Big(R - 4 \partial_{\mu}\phi \partial^{\mu}\phi - \frac{1}{12} \hat{H}_{\mu \nu \rho} \hat H^{\mu \nu \rho} - \frac14 F_{\mu \nu i} F^{\mu \nu i} \Big) \, ,
\label{alphasugra}
\eea
with 
\bea
\hat H_{ \mu\nu\rho}&=&3\left[\partial_{[\mu}b_{\nu\rho]} - \left(A_{[\mu}^i\partial_\nu A_{\rho]i} - \frac{1}{3}f_{ijk}A_{\mu}^iA_{\nu}^jA_{\rho}^k\right) \right] \\
F_{\mu\nu}^i&=&2\partial_{[\mu}A^i_{\nu]}-f^i{}_{jk}A_{\mu}^j A_{\nu}^k \, .
\eea
In this section we show how the double Yang-Mills formulation of DFT can reproduce the standard transformation rules and action of the heterotic supergravity. In the next one we inspect perturbative aspects of this theory using the relaxed form of the GKSA.

\section{Double Yang-Mills and the relaxed GKSA}
\label{RGKSA}
The GKSA was originally formulated in \cite{KL} and it has been explored in heterotic DFT \cite{HetKL} \cite{EA1} \cite{EA2}, Kaluza-Klein DFT \cite{KK}, Exceptional Field Theory \cite{EFT}, among others \cite{Relaxed} \cite{Currents}. In this part of the work we extend this ansatz to the double Yang-Mills formulation of heterotic field theory. We use a relaxed form of the GKSA, firstly introduced in \cite{Relaxed}.  

\subsection{Generalized metric formulation}
We start by considering a generalized metric ${H}_{M N}$, a generalized dilaton $d$ and a generalized gauged field $A_{M \a}$. The generalized metric satisfies a relaxed Kerr-Schild ansatz \cite{Relaxed} while the gauge field is linearly perturbed and the generalized dilaton has an arbitrary perturbation, i.e.,
\bea
\label{metricrelax}
{H}_{M N} & = & {H}_{o M N} + \kappa (\ov K_{M} K_{N} + K_{M} \ov K_{N}) + \frac{\kappa^2}{2} \ov{K}^2 K_{M} K_{N} \, , \\
{C}_{M \a} & = & {C}_{o M \a} - \kappa K_{M} J_{\a} \, , \\
d & = & d_{o} + \kappa f \, ,
\label{Cpert}
\eea
where ${H}_{o M N}$, ${C}_{o M N}$ and $d_{o}$ are generalized backgrounds,$f = \sum_{n=0}^{\infty} f^{(n)} \kappa^n$, $K_{M}$ and $\bar{K}_{M}$ are projected vectors,   
\bea
\bar{K}_{M} & = & \frac12({\eta}_{ M N} + { H}_{ M N}) \bar{K}^{ N} = \bar{P}_{ M N} \bar{K}^{ N} \, , \\ 
K_{ M} & = & \frac12({\eta}_{ M N} - { H}_{ M N}) K^{ N} = {P}_{ M N} {K}^{ N} \, ,
\eea
and  $\kappa$ is an order parameter. The null condition is given only on $K_{M}$,
\be
 \eta^{ M N} K_{ M} K_{ N} = 0 \, ,
\label{nullHdft}
\ee
since $\ov K_{M}$ is related to $J_{\a}$ through
\be
\eta^{ M N} \ov K_{ M} \ov K_{ N} = - J_{\a} J_{\b} \kappa^{\a \b} \, .
\ee

The parametrization of the ansatz is the same as (\ref{paramsugra})-(\ref{paramsugra2}) but the inverse metric receives a second order perturbation,
\bea
g^{\mu \nu} & = & \tilde g_{o}^{\mu \nu} + \kappa l^{(\mu} \bar{l}^{\nu)} + \frac{\kappa^2}{4} \bar l^2 l^{\mu} l^{\nu} \, .
\eea

Let us now analyze the pure Abelian Yang-Mills theory at the DFT level. We start with
\bea
S & = & \int d^{2D}X e^{-2d} H^{M N} F^{K}{}_{M \a} F_{K N}{}^{\a}   \\   & = & \int d^{2D}X e^{-2d} (H^{M N} \partial_{M} A^{K \a} \partial_{N} A_{K \a} - 2 H^{M N} \partial_{M} A^{K \a} \partial_{K} A_{N \a}) \, ,
\eea
where $d$ and $H^{M N}$ are constants and the only dynamical field is $A_{M \a}$. The equation of motion of the later is given by
\bea
H^{M N} \partial_{M}F_{K N} + 2 H^{M}{}_{K} \partial^{N}F_{M N} = 0 \, ,
\label{Max}
\eea
which can be understood as a duality invariant generalization of the Maxwell equation. Now we use the generalized Kerr-Schild ansatz on $A_{M \a}$ as
\bea
A_{M \a} = A_{o M \a} + \kappa K_{M} J_{\alpha} \, ,
\eea
in agreement with (\ref{Cpert}). The dynamics turns into
\bea
&& H^{MN} \partial_{M}(\partial_{K}(A_{o N}{}^{\alpha} + \kappa K_{N} J^{\alpha}) - \partial_{N}(A_{o K}{}^{\a} + \kappa K_{K} J^{\a})) \nn \\ &&
+ H^{M}{}_{K} \partial_{M}\partial_{N}(A_{o}^{N \a} + \kappa K^{N} J^{\alpha}) = 0 \, .
\eea
 To make contact with supergravity we parametrize the background fields as
\bea
{H}_{oMN} &=& \left(\begin{matrix}\eta^{\mu \nu} & 0 \\ 0 & \eta_{\mu \nu}  \end{matrix}\right)\ , \ \ \ \ \ \ \ \  \\
{C}_{o M}{}^\alpha &=& \frac 1 2 \left(\begin{matrix} - \eta^{\mu \rho} A_{o\rho}{}^i \\  A_{o\mu}{}^i\end{matrix}\right)\ ,
\eea
and the perturbations as
\bea
{K}_{ M} = \, \frac{1}{\sqrt{2}} \left( \begin{matrix} l^{\mu} \\ - l_{\mu}  \end{matrix} \right) \, , \quad
\ov{ K}_{ M} = \, \frac{1}{\sqrt{2}} \left( \begin{matrix} {\bar l}^{\mu} \\  {\bar l}_{\mu}  \end{matrix} \right) \,, \quad J_{\alpha}= \frac{1}{\sqrt{2}} j_{i} \, .  
\eea
The parametrization of the pure Abelian Yang-mills dynamics gives rise to the following first order contributions
\bea
-\frac{\kappa}{2} \eta^{\mu \nu} \partial_{\mu \rho}(l_{\nu} j^{i}) + \frac{\kappa}{2} \eta^{\mu \nu} \partial_{\mu \nu}(l_{\rho} j^{i}) = 0 \, .
\eea
In the next section we inspect the form of the relaxed generalized Kerr-Schild ansatz in the generalized frame formulation of DFT. As previously described in \cite{Tension}, we discussed about the equivalences of both formulations.

\subsection{Generalized frame formulation}

The relaxed generalized Kerr-Schild ansatz for the double Yang-Mills fields in the generalized frame formulation is given by,
\bea
\label{frame1}
{ E}_{ M}{}^{\ov{ A}} & = & {E}_{o M}{}^{\ov{ A}} + \frac12 \kappa { E}_{o N}{}^{\ov{ A}} { K}_{ M} \ov{ K}^{ N} \, , \\ \label{frame2} { E}_{ M}{}^{\underline{ A}} & = & { E}_{o M}{}^{\underline { A}} - \frac12 \kappa {E}_{oN}{}^{\underline{ A}} \ov{ K}_{ M} { K}^{ N}\, - \frac18 \kappa^2 \ov{K}^2 K_{M} K_{N} E_{o}^{N}{}_{\un A} \, , \label{GKSA} 
\eea
while the generalized gauge field and the generalized dilaton are perturbed as in the generalized metric formalism. The second order perturbation of $E_{M}{}^{\un A}$ has its coefficient fixed through
\bea
E_{M \un A} E^{M}{}_{\un B} + E_{M \ov A} E^{M}{}_{\ov B} = \eta_{A B} \, , \\
- E_{M \un A} E^{M}{}_{\un B} + E_{M \ov A} E^{M}{}_{\ov B} = H_{A B}
\eea
since the RHS of the previous expressions must remain invariant. In this formulation, the generalized metric is an element of $O(D,D)$ only if $\ov{K}^2=0$. In other words, the perturbation (\ref{metricrelax}) is not compatible with (\ref{GKSA}), which explains why the gauge field cannot be perturbed in the ordinary $O(D,D+K)$ framework. This formulation still contains an obstruction, but it appears before parametrization. This is very reasonable since the double Yang-Mills formulation of heterotic DFT written in terms of $O(D,D)$ multiplets is equivalent to the standard formulation written in terms of $O(D,D+K)$ multiplets.   

\subsection{Field redefinitions}
One interesting aspect of the different formulations of DFT is the need of field redefinitions to match with standard supergravity scenarios. This situation usually appears when the generalized metric has a non-trivial symmetry invariance because it encodes the b-field. The most canonical example is the Green-Schwarz mechanism. If we analyze the equation (\ref{redef}), this redefinition is mandatory since the metric $\tilde g$ transforms non-covariantly under gauge transformations according to (\ref{GSH}). When one works at the perturbative level, background objects could encode perturbations. 

Let us focus in the ordinary Kerr-Schild ansatz \cite{KS} for a generic $\tilde g_{\mu \nu}$,
\bea
\tilde g_{\mu \nu} = \tilde g_{o \mu \nu} + \kappa \tilde l_{\mu} \tilde l_{\nu} \, .
\eea
In order to have agreement at the perturbative level we are forced to impose a vector redefinition
\bea
\tilde l_{\mu} = l_{\mu} + A_{o \mu}{}^{i} j_{i} \, ,
\eea
and a field redefinition for the backgrounds 
\bea
\tilde g_{o \mu \nu} = g_{o \mu \nu} + \frac{\kappa^2}{2} l_{\mu} l_{\nu} j^2 \, . 
\eea
Now it is very clear that the background $\tilde g$ encodes a (second order) perturbation in terms of the ordinary $l$ vector. Moreover, the null and geodesic equations for the tilde fields impose new conditions. These relations were discussed in \cite{EA2} for Lorentz Green-Schwarz contributions in the generalized Kerr-Schild ansatz, and they are equivalent to the gauge Green-Schwarz contributions when the Lorentz connection is replaced by the gauge connection.

\section{Application}
\label{DCsec}
\subsection{Classical double copy at the DFT level}
The main results of this work allow us the inspection of the single and zeroth copy at the DFT level. The foundations of this relation and DFT can be consulted in \cite{OHDC}.  

Let us focus on the ungauged part of the equation of motion of the generalized metric with vanishing dilaton,
\bea
 {\cal R}_{ M  N} = P_{ M}{}^{ P} {\cal K}_{ P  Q} {\bar P}^{ Q}{}_{ N} +  {\bar P}_{ M}{}^{ P} {\cal K}_{ P  Q} P^{ Q}{}_{ N} = 0 \, ,
\eea
where
\bea
 {\cal K}_{ M  N} & = & \frac{1}{8} \partial_{ M} {H}^{ K  L} \partial_{ N} {H}_{ K  L} - \frac14 \partial_{ L} {H}^{ L  K} \partial_{ K}{H}_{ M  N} - \frac14 {H}^{ L  K} \partial_{K}\partial_{L}{H}_{ M  N} \nn \\ && - \frac12 \partial_{( M} {H}^{ K  L} \partial_{ L} {H}_{ N)  K} + \frac12 \partial_{ L} {H}^{ K  L} \partial_{( M} {H}_{ N)  K} + \frac12 \partial_{ L} {H}^{ K}{}_{( M} \partial_{ K} {H}^{ L}{}_{ N)} \nn \\ && + \frac12 {H}^{ K  L} \partial_{L( M} {H}_{ N)  K} + \frac12 {H}^{ K}{}_{( M} \partial_{K} \partial_{L} {H}^{ L}{}_{ N)} \, . \nn 
\eea
Now we can impose the ordinary generalized Kerr-Schild ansatz considering a flat ${H}_{o M N}$ background,
\bea
{H}_{M N} & = & {H}_{o M N} + \kappa (\ov K_{M} K_{N} + K_{M} \ov K_{N}) \, .
\eea
In this case it is useful to invoke the generalized version of the geodesic condition \cite{KL}
\bea
{\ov {K}}^{P} \partial_{P} {K}^{M} =  {K}^{P} \partial_{P}{\ov { K}}^{M} & = & 0 \, .
\eea
After imposing the previous conditions, the linear contributions reads
\bea
&& - \frac12 { H}_{o}^{K L} \partial_{K}(\partial_{L} K_{M} \ov{K}_{N} + K_{M} \partial_{L} \ov{K}_{N}) \nn \\ && + \ov { P}_{o M}{}^{K} \partial_{K} ( \partial_{L} K^{L} \ov{K}_{N})
- {P}_{o N}{}^{K} \partial_{K} (K_{M} \partial_{L}\ov{K}^{L})= 0 \, ,
\eea
while the quadratic contributions are null since they depend on the perturbation of the generalized dilaton. Now we define a generalized killing vector, $\xi_{M}$, in the following way
\bea
{\cal L}_{\xi} { H}_{M N}=0 \, ,
\eea
where ${\cal L}_{\xi}$ is the generalized Lie derivative. Furthermore, we choose a set of double coordinates such as $\xi=const.$ and
\bea
\xi^{M} K_{M} = \xi^{M} \ov K_{M} = 1 \, .
\eea
If we contract $\xi^{M} R_{M N}$ we obtain,
\bea
&& - \frac12 { H}_{o}^{K L} \partial_{K} \partial_{L} \ov{K}_{N} + \xi^{M} \ov { P}_{o M}{}^{K} \partial_{K} ( \partial_{L} K^{L} \ov{K}_{N})
- {P}_{o N}{}^{K} \partial_{K} \partial_{L} \ov{K}^{L}= 0 \, ,
\label{single1}
\eea
while contracting $\xi^{N} R_{M N}$ gives
\bea
&& - \frac12 { H}_{o}^{K L} \partial_{K} \partial_{L} {K}_{M} - \xi^{N} { P}_{o N}{}^{K} \partial_{K} ( \partial_{L} \ov K^{L} {K}_{M})
+ \ov{P}_{o M}{}^{K} \partial_{K} \partial_{L} {K}^{L} = 0 \, .
\label{single2}
\eea
Now we identify each null vector with a generalization of a $U(1)$ vector field, $A_{M}=A_{\un M}$ and $\ov A_{M} = \ov {A}_{\ov M}$ and we obtain the generalization of the single copy expression at the DFT level,
\bea
- \frac12 { H}_{o}^{K L} \partial_{K} \partial_{L} \ov{A}_{N} + \xi^{M} \ov { P}_{o M}{}^{K} \partial_{K} ( \partial_{L} A^{L} \ov{A}_{N})
- {P}_{o N}{}^{K} \partial_{K} \partial_{L} \ov{A}^{L} & = & 0 \, , \nn \\
- \frac12 { H}_{o}^{K L} \partial_{K} \partial_{L} {A}_{M} - \xi^{N} { P}_{o N}{}^{K} \partial_{K} ( \partial_{L} \ov A^{L} {A}_{M})
+ \ov{P}_{o M}{}^{K} \partial_{K}\partial_{L} {A}^{L} & = & 0 \, .
\label{DC}
\eea
Even when the previous expressions do not match exactly with the generalization of the Maxwell equation at the DFT level, it is straightforward to show that they embedded a pair of Maxwell equations at the supergravity level. We start by considering a null B-field at the supergravity level and then $\xi^{M}=(0,\xi^{\mu})$. The generalized metric is parametrized as
\bea
{H}_{MN} &=& \left(\begin{matrix} \eta^{\mu \nu} & 0 \\ 0 & \eta_{\mu \nu}  \end{matrix}\right)\ ,
\eea
while the pair of generalized gauge vectors reproduces the ordinary Abelian fields $A_{M} \rightarrow A_{\mu}$ and $\ov A_{M} \rightarrow \ov A_{\mu}$. The geodesic condition turns into
\bea
A_{\mu} \partial^{\mu} \ov A_{\nu} & = & 0 \, , \nn \\
\ov A_{\mu} \partial^{\mu} A_{\nu} & = & 0 \, \nn ,
\eea
while the ordinary killing vector satisfies
\bea
\xi^{\mu} \partial_{\mu}(A_{\nu} \ov A_{\rho}) = 0 \, .
\eea
The parametrization of (\ref{DC}) is given by
\bea
\Box A_{\nu} - \partial_{\nu}(\partial^{\rho} A_{\rho}) & = & 0 \leftrightarrow \partial_{\mu}F^{\mu \nu} = 0 \\
\Box A_{\nu} - \partial_{\nu}(\partial^{\rho} \ov A_{\rho}) & = & 0 \leftrightarrow \partial^{\mu} \ov F_{\mu \nu} = 0 \, ,
\eea
where we have defined the curvatures of the Abelian gauge fields in the usual way
\bea
F_{\mu \nu} & = & 2 \partial_{[\mu} A_{\nu]} \, , \nn \\
\ov F_{\mu \nu} & = & 2 \partial_{[\mu} \ov A_{\nu]} \, \nn . 
\eea

On the The zeroth copy relation can be easily obtained contracting an extra generalized Killing vector $\xi$ in (\ref{single1}) or (\ref{single2}), after introducing the scalar function $\varphi$ using the redefinition $\kappa_{\textrm{DFT}} \rightarrow \kappa \varphi$. The resulting equation is given by
\bea
- \frac12 { H}_{o}^{K L} \partial_{K} \partial_{L} \varphi + \xi^{M} \ov { P}_{o M}{}^{K} \partial_{K} \partial_{L} (\varphi K^{L})
- \xi^{N} {P}_{o N}{}^{K} \partial_{K L}(\varphi \ov{K}^{L})= 0 \, ,
\eea
which embeds the standard supergravity zeroth copy relation,
\bea
\Box \varphi = 0 \, .
\eea

\section{Discussion}
\label{Dis}
The double Yang-Mills formulation of heterotic DFT forces the apparition of a generalized gauge field/symmetry in the double geometry. Moreover, the $O(D,D)$ generalized metric transforms according to a generalized Green-Schwarz mechanism, which is inherited to the b-field and also to the metric tensor. For this reason, a field redefinition is mandatory to remove this last transformation as we show in (\ref{redef}). This formulation explains the agreement between \cite{HetDFT} and the proposal given in \cite{PK}, where a pure double Yang-Mills formulation was proposed. 

The extra terms (beyond $F^2$ contributions) which we find here are directly related to Chern-Simons contributions at the supergravity level. This kind of terms cannot be written in the double geometry \footnote{We thank K. Lee for this observation.} since the notion of the exterior derivative is not fully understood in the present formulation of DFT. In heterotic supergravity, the need of Chern-Simons terms is related to the lack of invariance of the square of the curvature of the b-field under gauge transformations. Following this logic, a more general notion of Chern-Simons terms in the double geometry could involve particular combinations of the $O(D,D)$ fundamental fields, i.e. $CS_{MNP}$, which might be used in order to rewrite the double Yang-Mills action in a compact form, curing the lack of invariance of the pure double Yang-Mills term.

On the other hand, here we revisit the tension between the generalized metric/frame formalism of DFT when one considers the GKSA. Particularly, using the standard form of the GKSA we show that the D-dimensional gauge field, $A_{\mu i}$, cannot be perturbed in the generalized frame formulation. Furthermore, we relax the conditions of the GKSA mimicking \cite{Relaxed} in order to perturb this field at the DFT level. As happens in the ordinary case, the obstruction is still present since the generalized metric is not an $O(D,D)$ element unless $J_{\alpha}=0$. 

It would be natural to propose an ordinary GKSA over the $O(D,D+K)$ invariant DFT and then impose a suitable splitting as
\bea
\ov{\cal K}_{\cal M} = (\ov K_{M}, J_{\a}) \, , \quad \quad {\cal K}_{\cal M} = (K_{M}, 0) \, .
\eea
Unfortunately, this proposal also leads to $J_{\a}=0$ according to the gauge fixing of the $O(D,D+K)$ generalized frame. 

Finally, using the main results of this paper we explore the classical double copy relation at the DFT level. A very important observation of equation (\ref{DC}) is the fact that after the identification of the generalized null vectors with Abelian gauge fields in the EOM of the generalized metric contracted with a generalized Killing vector, the form of the equation is not exactly a Maxwell-like equation as (\ref{Max}) as happens at the supergravity level. However, the agreement occurs upon parametrization, as expected.

\section{Summary}
\label{Conclu}

We present a double Yang-Mills formulation of heterotic DFT, where the fundamental fields are in representations of $O(D,D)$ and a generalized gauged field/symmetry explicitly appears in the double geometry. The formalism is equivalent to the standard formulation of heterotic DFT and contains a pure double Yang-Mills theory in its Lagrangian formulation as proposed in \cite{PK}. We use this framework to explore a relaxed version of the GKSA. The generalized background metric contains up to quadratic perturbations generated by a single null vector, while the generalized gauge field is linearly perturbed before parametrization. As an application we explore the classical double copy correspondence at the DFT level and we find that after obtaining the generalized version of the single copy, the form of the equation does not match with the generalization of the Maxwell equation and the agreement occurs upon parametrization.    

\section*{Acknowledgements}
We are very grateful to K. Lee for e-mail correspondence and enlighten comments. We also thank to the organisers of the school ``Integrability, Dualities and Deformations" which ran from 23 to 27 August, 2021, in Santiago de Compostela and virtually, since they provided discussion sessions where this project began. Finally, E.L would like to thank CONICET for supporting his work. The work of SRC is partially supported by grants from the Infosys Foundation to CMI.

\appendix

\section{Relaxed GKSA and the DFT Lagrangian}
If we focus on the ungauged part of the double Yang-Mills Lagrangian assuming constant backgrounds and  with vanishing dilaton, i.e.,
\bea
\frac{1}{8} {H}^{MN} \partial_M {H}^{KL}  \partial_N {H}_{KL} -  \frac{1}{2} {H}^{MN} \partial_N {H}^{KL}  \partial_L {H}_{MK} \, ,
\eea
it is straightforward to verify that the sixth and fifth order contributions vanish since $K_{M}$ is a null vector. The $\kappa^4$ contributions are
\bea
&& - \frac18 K^{M} K^{N} {\ov K}^4 \partial_{M}{K_{Q}} \partial_{N}{K^{Q}} - \frac18 K^{M} K^{N} {\ov K}_{P} {\ov K}^{Q} {\ov K}^2 \partial_{M}{K_{Q}} \partial_{N}{K^{P}} \nn \\ && + \frac18 K^{M} K^{N} {\ov K}^{P} {\ov K}^{Q} {\ov K}^2 \partial_{M}{K_{P}} \partial_{N}{K_{Q}} - \frac18 H_{o}^{M N} K_{M} K^{P} {\ov K}^4 \partial_{P}{K_{Q}} \partial_{N}{K^{Q}} \, ,
\eea
the $\kappa^3$ contributions are
\bea
&& - \frac14 K^{M} {\ov K}^2 {\ov K}^{P} \partial_{P}{K_{Q}} \partial_{M}{K^{Q}} - \frac14 K^{M} {\ov K}_{N} {\ov K}^{P} {\ov K}^{Q} \partial_{P}{K_{Q}} \partial_{M}{K^{N}} \nn \\ && + \frac14 K^{M} {\ov K}^2 {\ov K}^{P} \partial_{M}{K_{Q}} \partial_{P}{K^{Q}} + \frac14 K^{M} {\ov K}_{N} {\ov K}^{P} {\ov K}^{Q} \partial_{M}{K_{P}} \partial_{Q}{K^{N}} \nn \\ && - \frac14 H_{o}^{M N} K_{M} {\ov K}^{P} {\ov K}^2 \partial_{P}{K_{Q}} \partial_{N}{K^{Q}} + \frac14 H_{o}^{M N} K_{M} K^{P} {\ov K}^{Q} \partial_{N}{K_{Q}} \partial_{P}{{\ov K}^2} \nn \\ && + \frac14 H_{o}^{M N} K^{P} {\ov K}^{Q} {\ov K}^2 \partial_{M}{K_{Q}} \partial_{P}{K_{N}} - \frac14 H_{o}^{M N} K_{M} K^{P} {\ov K}^2 \partial_{P}{K_{Q}} \partial_{N}{{\ov K}^{Q}} \nn \\ && + \frac14 H_{o}^{M N} K_{M} K^{P} {\ov K}^{Q} \partial_{P}{K_{Q}} \partial_{N}{{\ov K}^2} - \frac14 H_{o}^{M N} K^{P} {\ov K}_{Q} {\ov K}^2 \partial_{P}{K_{M}} \partial_{N}{K^{Q}} \nn \\ && - \frac14 H_{o}^{M N} K_{M} K^{P} {\ov K}^2 \partial_{N}{K^{Q}} \partial_{P}{{\ov K}_{Q}} - \frac14 H_{o}^{M N} K^{P} {\ov K}_{M} {\ov K}^2 \partial_{P}{K_{Q}} \partial_{N}{K^{Q}} \nn \\ && - K^{M} {\ov K}^2 {\ov K}^{P} \partial_{Q}{K_{P}} \partial_{M}{K^{Q}} - \frac12 K^{M} K^{N} {\ov K}^2 \partial_{M}{K_{Q}} \partial_{N}{{\ov K}^{Q}} \, ,
\eea
and, finally, the $\kappa^2$ contributions are
\bea
&& \frac14 H_{o}^{M N} {\ov K}^2 \partial_{M}{K_{Q}} \partial_{N}{K^{Q}} + \frac14 H_{o}^{M N} {\ov K}_{P} {\ov K}^{Q} \partial_{M}{K_{Q}} \partial_{N}{K^{P}} \nn \\ && + \frac14 H_{o}^{M N} {\ov K}^{P} {\ov K}^{Q} \partial_{M}{K_{P}} \partial_{N}{K_{Q}} - \frac12 H_{o}^{M N} {\ov K}_{P} {\ov K}^{Q} \partial_{Q}{K_{M}} \partial_{N}{K^{P}} \nn \\ && - \frac12 H_{o}^{M N} K_{M} {\ov K}^{P} \partial_{N}{K^{Q}} \partial_{P}{{\ov K}_{Q}} + \frac12 H_{o}^{M N} K_{M} {\ov K}^{P} \partial_{Q}{K_{P}} \partial_{N}{{\ov K}^{Q}} \nn \\ && - \frac12 H_{o}^{M N} {\ov K}_{M} {\ov K}^{P} \partial_{P}{K_{Q}} \partial_{N}{K^{Q}} - \frac12 H_{o}^{M N} K^{P} {\ov K}^{Q} \partial_{P}{K_{M}} \partial_{N}{{\ov K}_{Q}} \nn \\ && - \frac12 H_{o}^{M N} K_{M} K^{P} \partial_{P}{{\ov K}_{Q}} \partial_{N}{{\ov K}^{Q}} - \frac12 H_{o}^{M N} {\ov K}^2 \partial_{Q}{K_{M}} \partial_{N}{K^{Q}} \nn \\ && - \frac12 H_{o}^{M N} K_{M} {\ov K}^{P} \partial_{N}{K^{Q}} \partial_{Q}{{\ov K}_{P}} + \frac12 H_{o}^{M N} K^{P} {\ov K}^{Q} \partial_{M}{K_{Q}} \partial_{P}{{\ov K}_{N}} \nn \\ && - \frac12 H_{o}^{M N} K^{P} {\ov K}_{M} \partial_{P}{K_{Q}} \partial_{N}{{\ov K}^{Q}} - \frac12 H_{o}^{M N} {\ov K}_{M} {\ov K}^{P} \partial_{Q}{K_{P}} \partial_{N}{K^{Q}} \, .
\eea
On the other hand, the generalized dilaton contributions are given by
\begin{eqnarray} 
\mathcal{L}_{d} & = & 4 H^{MN} \partial_M d  \partial_N d - 2 \partial_M H^{MN} \partial_N d  \nonumber\\
& = & 4 \kappa^2 \Bigg[H_{o}^{MN} + \kappa \Big(\bar{K}^M K^N + K^M \bar{K}^N\Big)\Bigg] \partial_M f \partial_N f \nonumber\\
&& - 2 \kappa^2 \partial_M \Bigg[ \Big(\bar{K}^M K^N + K^M \bar{K}^N\Big)\Bigg]  \partial_N  f \ . 
\end{eqnarray}

The gauged part can be easily computed using the linear ansatz for the C-field,
\bea
\Delta_{\a \b}-\Delta_{o\a \b} & = & -2 \kappa K^{M} J_{(\a} C_{oM \b)} \\ 
\Theta_{M N}- \Theta_{o M N} & = & -2 \kappa K_{(M} J^{\a} C_{oN)\a} + \kappa^2 K_{M} K_{N} J^2 \, .
\eea
The first order contributions can be easily computed using (\ref{LagC}) since $\Delta^{-1}=\Delta^{-1}_{o} + \mathcal{O}(\kappa^2)$. Then, the $\kappa$ contribution for the gauge sector is,
\bea
  4 f_{\beta\gamma}^{\alpha} C^{M \b}  C^{N \gamma} \partial_M (K_{N} J_{\a})  
\eea
Higher-order contributions can be straightforwardly computed following the same logic to any desired order.

\end{document}